\documentclass{article}
\usepackage[utf8]{inputenc}
\usepackage{amsmath}
\usepackage{graphicx}
\usepackage{hyperref}
\usepackage[left=2.5cm,right=2.5cm,
    top=2cm,bottom=3cm,bindingoffset=0cm]{geometry}

\title{Numerical Analysis of Antenna Parameter Influence on Brightness Temperature in Medical Microwave Radiometers}
\author{Maxim V. Polyakov$^*$ and Danila S. Sirotin\footnote{Volgograd State University, Universitetsky pr., 100, Volgograd, Russia}
\\m.v.polyakov@volsu.ru, d.sirotin@volsu.ru
}
\date{January 17, 2025}

\begin{document}

\maketitle



\begin{abstract}
\noindent This article presents a study on the influence of antenna parameters in medical microwave radiometers on brightness temperature. A series of computational experiments was conducted to analyse the dependence of brightness temperature on antenna characteristics. Various antenna parameters and their effect on the distribution of electromagnetic fields in biological tissues were examined. It was demonstrated that considering the antenna mismatch parameter is crucial when modelling the brightness temperature of biological tissues, contributing about 2 percent to its formation. The depth range of brightness temperature measurement was determined. The dependence of brightness temperature on the antenna diameter and frequency was established. The findings of this study can be applied to improve medical microwave radiometers and enhance their efficiency in the early diagnosis of various diseases.

\vspace{2mm}
\noindent \textbf{Keywords:} numerical modelling, brightness temperature, microwave radiothermometry, antenna mismatch
\end{abstract}

\section{Introduction} \label{intro}

A microwave radiometer is a device that uses microwave radiation to measure the temperature of \mbox{objects \cite{Filatov2015, Vesnin2021}.} It is based on the principle that the microwave radiation of an object correlates with its temperature, as described by the Stefan-Boltzmann law. The device measures the natural microwave radiation emitted by the object to determine its temperature. Microwave radiometers are widely used, including in medicine, because their measurements are non-invasive and do not harm patients \cite{Goryanin2020, Shevelev2022, Vesnin2017}.

Microwave radiothermometry is used in medicine to measure the temperature of tissues and internal organs within the human body \cite{Sidorov2021, Shah2011}. This technology finds applications in various medical fields, such as oncology, neurosurgery, and patient rehabilitation \cite{Shevelev2022, Losev2017, Kolesov1993, Petrova2022}.
In oncology, microwave radiothermometry is used to monitor tumour temperature during hyperthermic therapy \cite{Dubois1993}. Hyperthermic therapy is a cancer treatment approach that involves heating the tumour to a high temperature to destroy cancer cells \cite{Kang2020, Raouf2021}. A microwave radiometer enables the precise monitoring and regulation of tumour temperature, minimising damage to surrounding healthy tissues.
In neurosurgery, microwave radiothermometry is used to measure brain temperature during surgical procedures. This enables surgeons to monitor and prevent damage to brain tissue by ensuring the maintenance of an optimal temperature \cite{Rodrigues2018}.
In diagnosis and rehabilitation, microwave radiothermometry can be used to measure the temperature of muscles and joints during physiotherapeutic procedures \cite{Laskari2023}. This enables physiotherapists to adjust the intensity and duration of treatments, ensuring optimal thermal effects on tissues and facilitating the recovery process.

The study \cite{Goryanin2020} offers a comprehensive description of passive microwave radiometry as a method for biomedical research. The authors emphasise its potential for diagnosing various diseases and monitoring physiological parameters of the body.
The article \cite{Stauffer2014} discusses the use of microwave radiometry for non-invasive monitoring of brain temperature. The results demonstrate the high accuracy of the method in clinical practice.
Millimetre-wave thermometry was investigated as a new approach to non-invasive temperature measurement and the high accuracy of the method was confirmed in various clinical scenarios \cite{Zhadobov2015}.
The \mbox{work \cite{Singh2024}} examines the prospects of using microwave radiothermometry for intracranial pressure monitoring and proposes effective algorithms for processing the acquired signals.

Another important application of microwave radiothermometry is the diagnosis of cancers such as breast cancer \cite{Khoperskov2022}. Temperature changes in tissues can indicate the presence of tumour processes, which can be used to detect and localise tumours based on their thermal characteristics \cite{Glazunov2021}.

An important issue when using microwave radiothermometry is the accuracy of measurements. The accuracy of microwave radiothermometry depends on various factors, including the quality of the equipment, measurement techniques, and the experience of the physician. In general, modern microwave radiometers offer high accuracy and can measure temperature with a precision of up to 0.1 $^\circ$C \cite{Polyakov2023}. However, accuracy may vary depending on the specific situation and measurement conditions. The main factors influencing measurement results are as follows:
\begin{itemize}
\item The radiation pattern of an antenna determines how the antenna receives electromagnetic waves from the environment. Different radiation pattern shapes can affect the antenna's ability to measure the brightness temperature of an object
\item The choice of frequency range for the antenna should match the radiometer's radiation frequency. Failure to match the frequency range may result in distorted measurements.
\item The quality of the coupling between the antenna and the radiometer affects the efficiency of signal transmission. Improper coupling can result in signal loss and reduced measurement accuracy. For best results, it is important to calibrate the equipment. \end{itemize}

Numerical modelling of antenna applicators is used to optimize their design and evaluate their \mbox{performance \cite{Rodrigues2014, Yazdandoost2021}.} This allows for improved efficiency and accuracy of measurements made with microwave radiometers. Numerical modelling enables the investigation of various parameters of antenna applicators, such as frequency range, shape, size, materials, and antenna placement \cite{Sedankin2020}. Modelling helps determine the optimal parameters that ensure maximum accuracy and reliability of measurements. Additionally, numerical modelling allows for the assessment of the impact of various factors, such as electromagnetic interference or interaction with the surrounding environment, on the performance of antenna applicators. This helps enhance measurement stability and reduce potential distortions in results.

The study of the influence of microwave radiometer antenna parameters on the brightness temperature is a relevant topic due to the growing interest in the development of more accurate and efficient methods for measuring the temperature of various objects, including biological tissues, food products, materials, etc. Optimising the parameters of radiometer antennas can significantly enhance measurement accuracy, improve resolution, and expand the scope of applications. Therefore, this topic is of great importance to the scientific community and engineering practice.

The main goal of this work is to study the influence of various antenna parameters in microwave radiometers on brightness temperature. The research aims to identify key characteristics that contribute to improving the accuracy and efficiency of brightness temperature measurements in the microwave range. The results of this study may have significant practical implications for the development of more accurate and reliable microwave radiometers.

\section{Methods}\label{MathematicalAndNumericalModel}

\subsection{Geometry of the studied model}\label{GeometryProblems}

\begin{figure}[h!]
    \begin{center}
    \includegraphics[scale=0.45]{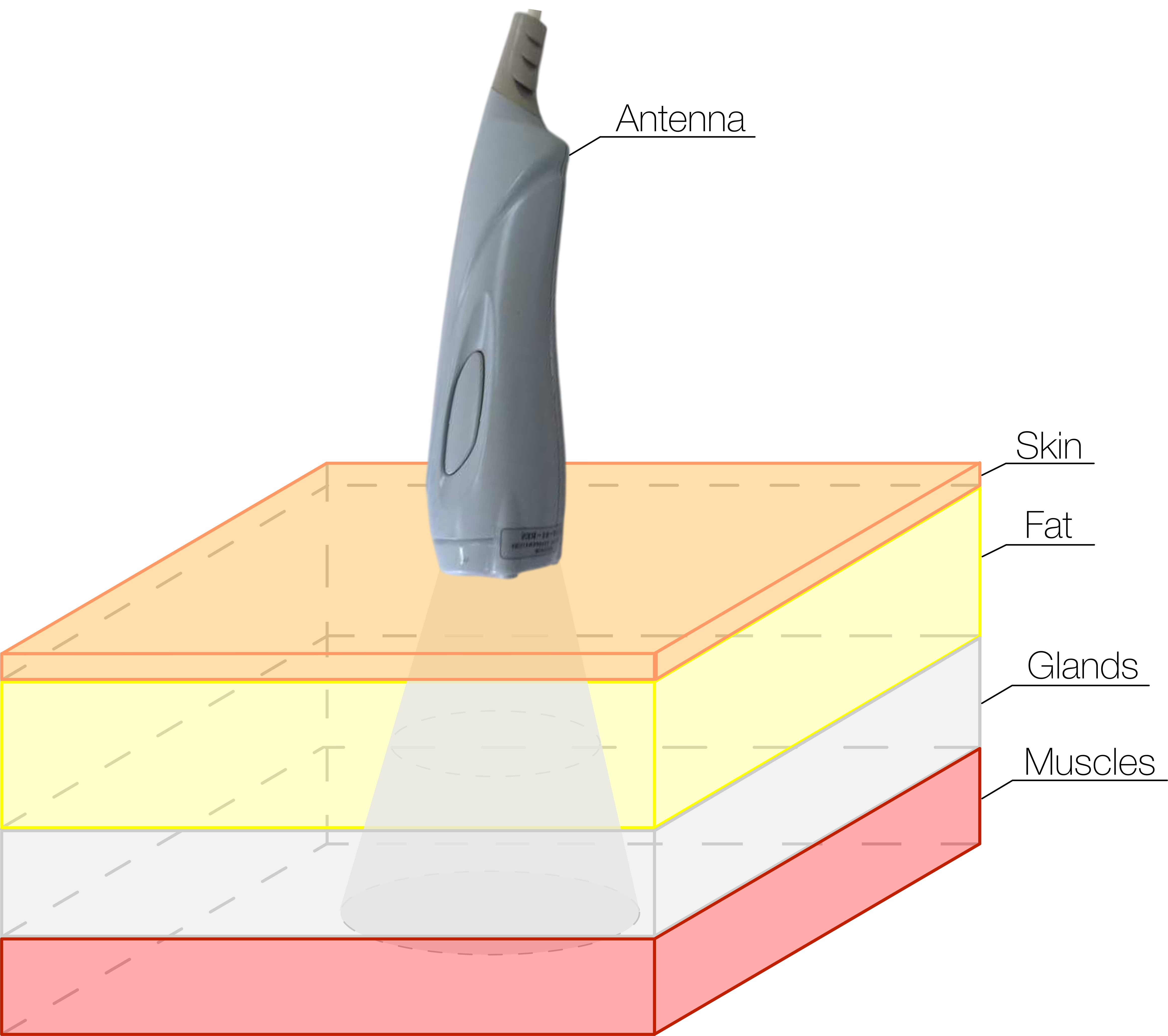}\\[2mm]
    \caption{Schematic representation of the multilayer structure of biological tissue with a microwave radiometer antenna} \label{fig1}
    \end{center}
\end{figure}

The multilayer structure of biological tissue represents a complex system composed of several layers, each characterised by distinct thermal and electromagnetic properties. The first layer is the skin, which is the outermost layer of biological tissue, in contact with the environment. This is followed by layers of subcutaneous fat, glandular tissue, and muscles. Each of these layers possesses a unique set of physical characteristics that influence the transmission and absorption of electromagnetic waves. In this work, we consider electrical conductivity, dielectric permittivity, and magnetic permeability. The antenna is placed on the surface of the biological tissue (Figure \ref{fig1}). Depending on the antenna design, it can operate at different frequencies and have a specific radiation pattern.

Further, the influence of various antenna characteristics on the brightness temperature it generates will be discussed.

\subsection{Mathematical and numerical modelling of brightness temperature in biological tissues}


We consider the electric field of the radiometer antenna in the approximation of a monochromatic plane wave, attenuating with depth. An antenna with a frequency of several GHz allows for measuring the thermal radiation from biological tissues within a specific frequency range $f_{\min} \le f \le f_{\max}$. Due to the multilayer structure, biological tissue has a non-uniform spatial temperature. This method provides an average weighted temperature $T_B$ for a certain internal region. The error of the radiothermometry method is also influenced by the receiver's noise temperature $T_{REC}$, the effects of antenna mismatch (determined by the coefficient $S_{11}(\omega)$) and environmental factors ($T_{EMI}$). As a result, the brightness temperature is determined by an integral representation of the form:

\begin{equation}\label{eq:2.1}
	T_B =\frac{1}{\Delta f} \int\limits_{f_{\min}}^{f_{\max}} \left\{ s_{11} \left[ T_{EMI} + \int\limits_{V_b} \Omega(x,y,z, f)\, T(x,y,z)\,dV \right] + |S_{11}(f)|^2 T_{REC}
	\right\}\,df \,,
\end{equation}
where $\Delta f=f_{max}-f_{min}$, $s_{11} = 1 - |S_{11}|^2$ accounts for the antenna mismatch. The integration is performed over a certain volume $V_b$, within which the radiation measured by the antenna is generated. 

The value
\begin{equation}\label{weight}
	\Omega=\frac{P_d(x,y,z;\omega)}{\int_{V_b} P_d dV}\,
\end{equation}
defines the weighting function with normalisation taken into account $\int_{V_b} \Omega\,dV=1$.

The expression for the electromagnetic energy density is equal to
\begin{equation}\label{Pd}
	P_d = \frac{1}{2}\sigma(x,y,z;\omega)\cdot |\vec{E}(x,y,z;\omega)|^2,
\end{equation}
where $\sigma$ is the electrical conductivity and $\vec{E}$ is the vector of the electric field generated by the antenna in the biological object.

The spatial distribution of the electric field in the monochromatic limit is described by the Helmholtz equation
\begin{eqnarray}\label{eq-Gelmgoltz}
	\Delta\, \vec{E}+\frac{\omega^2}{c^2}  \varepsilon \,\vec{E} = - \nabla \left( \vec{E}\cdot \vec{\nabla} (\ln\,\varepsilon) \right) \,,
\end{eqnarray}
where $\varepsilon(x,y,z;\nu)$ is the permittivity, $c$ is the speed of light in vacuum, $\omega=2\pi \nu$. 
The right side of \mbox{equation (\ref{eq-Gelmgoltz})} clearly shows the influence of the inhomogeneity of dielectric properties in a biological tissue. 

\begin{figure}[h!]
    \begin{center}
    \includegraphics[scale=0.45]{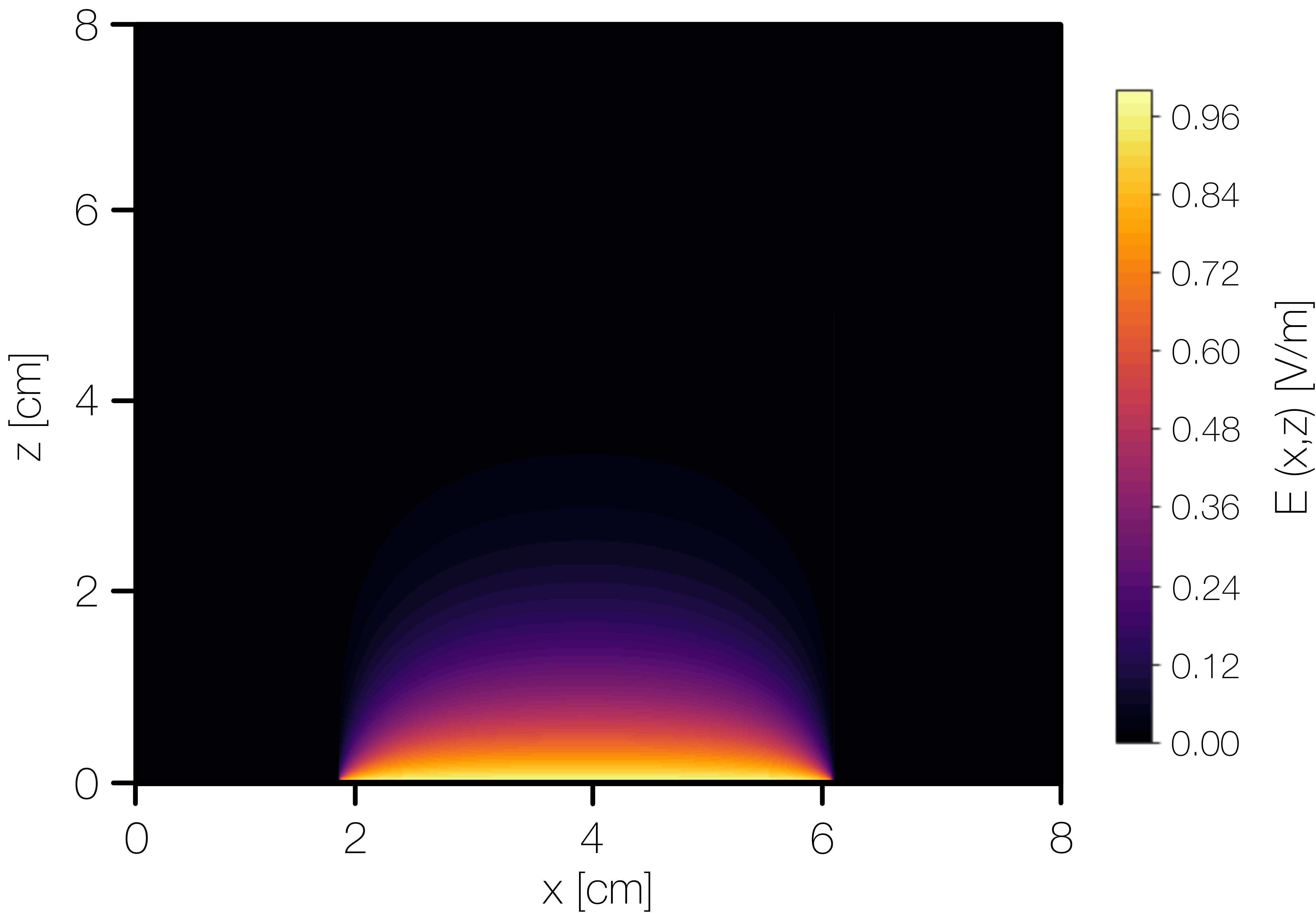}\\[2mm]
   \caption{An example of the calculated distribution of the electric field intensity $\vec{E}$ in the multilayer biological tissue in the plane beneath the antenna.} \label{fig2}
    \end{center}
\end{figure}

Figure \ref{fig2} illustrates the distribution of the electric field intensity in a multilayer biological tissue. The maximum intensity is observed on the surface of the tissue and gradually decreases with increasing penetration depth. Thus, the electric field attenuates as it passes through the various layers of the tissue.

Heat dynamics is determined by the heat conduction equation with different sources \cite{Khoperskov2022, Pennes-1948thermal-model} 
\begin{equation}\label{Main-model}
	\rho (\vec{r})\, C(\vec{r}) \frac{\partial  T(\vec{r},t)}{\partial t} = \nabla \left( \lambda(\vec{r}) \nabla T \right) +Q^{(met)}(\vec{r},t)+ Q^{(bl)}(\vec{r}) \,, 
\end{equation}
where $\rho$ is the mass density, $C$ is the  heat capacity of tissue, $T$ is the thermodynamic temperature, $\lambda$ is the thermal conductivity coefficient of biological tissue, $\vec{r} = \{x, y, z\}$, $\nabla$ is the nabla operator. 

The boundary conditions between the biological tissue (skin) and the environment are based on the continuity of the energy flux
\begin{equation} \label{boundary}
	\lambda(\vec{r})\,\left(\vec{n}\cdot\nabla T(\vec{r},t)\right) = h\cdot (T_{air}-T(x, y, z, t)),
\end{equation}
where $\vec{n}$ is the normal vector to the boundary of the interface ``biological tissue~-- environment'', $h$ is the heat transfer coefficient (W/m$^2 \cdot K$), $T_{air}$ is the ambient temperature.

The physical characteristics of biological tissues used in this study were taken from \cite{Khoperskov2022}.

\subsection{The method for calculating the antenna mismatch parameter}

\begin{figure}[h!]
    \begin{center}
    \includegraphics[scale=0.6]{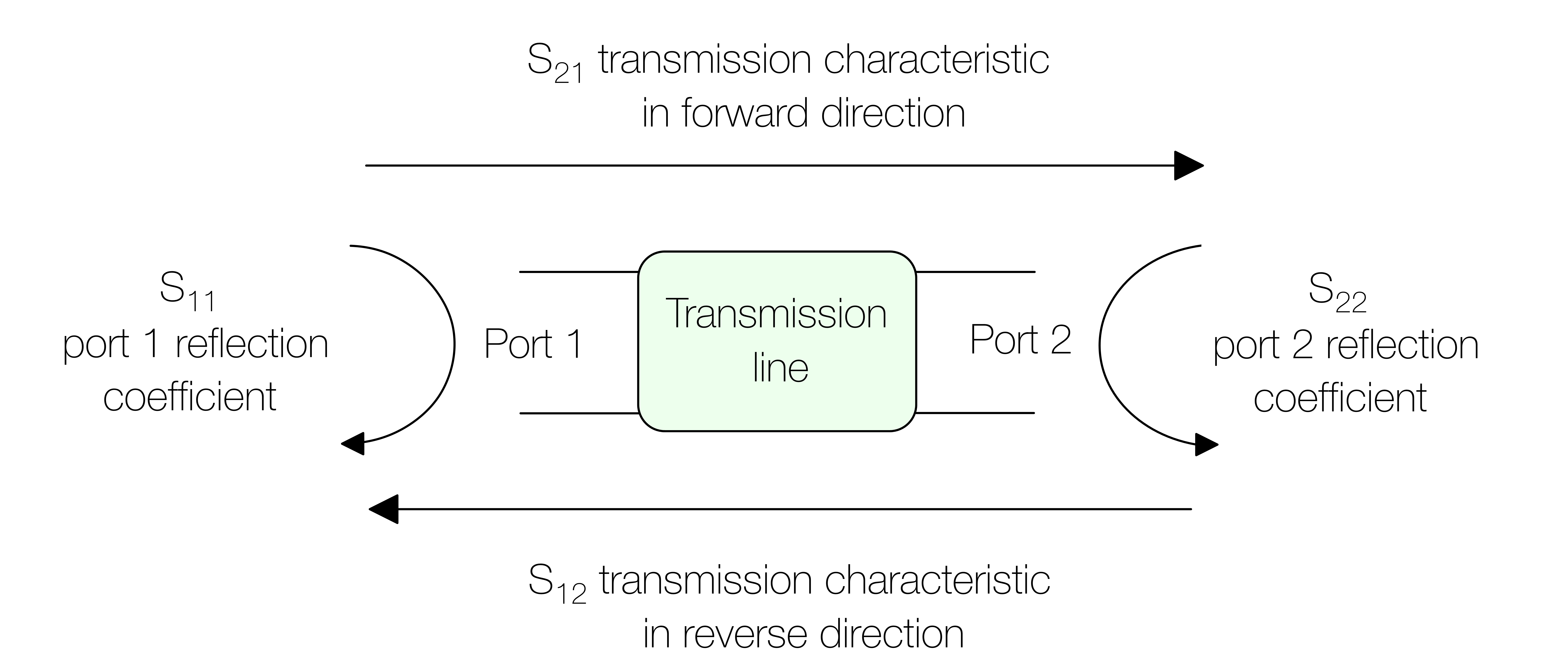}\\[2mm]
     \caption{Schematic representation of S-parameters for a microwave radiometer} \label{sparameters}
    \end{center}
\end{figure}

S-parameters are coefficients that describe the extent to which electromagnetic waves are reflected or transmitted through a device. They are used to characterise the signal transmission in radio frequency systems and antennas. Each element of the S-matrix corresponds to the relationship between each pair of inputs and outputs in the device. The general scheme of S-parameters for a transmission line is presented in Figure \ref{sparameters}.

The $S_{11}$ coefficient, also known as the reflection coefficient, is an element of the matrix that describes the portion of the input signal reflected back to the source. The number of elements in the S-matrix depends on the number of inputs and outputs in the device. This parameter, $S_{11}$, is utilised in the calculation of brightness temperature for the microwave radiothermometry method.

The value of parameter $S_{11}$ should be minimum at the operating frequency, provided the antenna is matched and properly tuned. The wave should not be reflected when power is supplied to the antenna. The magnitude of the parameter, $|S_{11}|$, is expressed in decibels. In the context of microwave device theory, it is referred to as return loss.

Return loss is defined as the negative value of the reflection coefficient. A high positive return loss indicates that the reflected power is small compared to the incident power, signifying good impedance matching between the transmission line and the load. The reflection coefficient parameter is defined as follows:

\begin{equation}
\label{eq:2.2}
 \Gamma=\frac{V_r}{V_i} ,
\end{equation}
where $V_r$ is the amplitude of the reflected wave, and $V_i$ is the amplitude of the incident wave.

The reflection coefficient is determined by the load impedance at the end of the transmission line and the characteristic impedance of the line.

Wave propagation coefficient
\begin{equation}
\label{eq:2.3}
\beta_{v}= \frac{2 \pi f l}{c}, 
\end{equation}
where $f$ is the signal frequency (Hz), $l$ is the wavelength, and $c$ is the speed of light ($\simeq 3 \cdot 10^{8}$ m/s).

Define the input impedance of the antenna
\begin{equation}
\label{eq:2.4}
Z_{in}=Z_{0}\frac{Z_{L}+Z_{0}\cdot \text{tan}(\beta_{v})}{Z_{0}+Z_{L}\cdot \text{tan}(\beta_{v})}, 
\end{equation}
where $Z_L$ is the total load impedance, and $Z_{0}$ is the input impedance of the antenna.

Reflection coefficient is
\begin{equation}
\label{eq:2.5}
S_{11}=\frac{Z_{in}-Z_{0}}{Z_{in}+Z_{0}}, 
\end{equation}
where $Z_{0}$ is the antenna impedance.

The standing wave ratio (SWR) parameter is the ratio of the maximum and minimum voltage amplitude values on the transmission line or antenna. It characterises the relationship between the waves reflected from the load and those incident upon it, and is determined by the following formula:

\begin{equation}
\label{eq:2.6}
\text{SWR}=\frac{1+|S_{11}|}{1-|S_{11}|}.
\end{equation}

If the SWR value is equal to one, it means that the wave is not reflected from the load and is fully incident upon it, which indicates that the load impedance matches the transmission line impedance, representing the optimal calculation result. Conversely, the higher the SWR value, the worse the impedance match, and the more waves are reflected from the load, which can lead to power loss and signal degradation.

The variation of the electric field from the antenna, centrally above the sphere, is calculated using the radiation pattern. The radius remains constant, while the angle parameters $\theta$ and $\phi$ are varied. The values are normalised so that the maximum equals one. This normalisation is achieved by orienting the radiation source along the $Z$-axis, resulting in the field component depending on the angle $\theta$

\begin{equation}
\label{eq:2.7}
 F(\theta, \phi)=\frac{E_{\theta}}{E_{\theta}(\max )} ,
\end{equation}
where $F(\theta, \phi)$ represents the field normalisation pattern, which depends on $\theta$ and $\phi$; $E_{\theta}(\max)$ is the maximum value of the field on the sphere.

The field pattern $F(\theta, \phi)$ does not contain imaginary or complex parts, and the value of $E_{\theta}$ takes positive values. This property defines linear polarisation in the direction of orientation $\theta$. The point where the phase is zero is determined, and the other phases are defined relative to this point. In this case, the amplitude is normalised to unity, allowing comparison of relative amplitude levels in different directions

\begin{equation}
\label{eq:2.8}
\begin{aligned}
A_{i} = 1,\\*
\varphi_{i} = e^{j\left(\beta x_{i}+\psi z_{i}\right)},
\end{aligned}
\end{equation}
where $A_{i}$ is the amplitude component for the $i$-th antenna element; $\varphi_{i}$ is the phase component for the $i$-th antenna element.

The parameter $\beta$ is one of the components of the phase factor and represents the angle between the maximum direction and the $x_{i}$ axis at each stage of the simulation, defined by the formula

\begin{equation}
\label{eq:2.9}
\begin{aligned}
\beta  = kd\sin(\theta)\cos(\phi),
\end{aligned}
\end{equation}
where $k=\frac{2\pi}{\lambda}$ is the angular wave number, $\lambda$ is the wavelength, and $d$ is the diameter of the antenna.

The parameter $\psi$ is also a component of the phase factor, defining the angle between the radiation pattern axis and the axis $z_{i}$ at each step of the simulation, and is determined by the formula

\begin{equation}
\label{eq:2.10}
\begin{aligned}
\psi  = kd\sin(\theta)\sin(\phi).
\end{aligned}
\end{equation}

The resultant electric field for the computational domain is
\begin{equation}
\label{eq:2.11}
\begin{aligned}
 E(\theta, \phi)=\sum_{i=1}^{n} A_{i} \varphi_{i} e^{j\left(\beta x_{i}+\psi z_{i}\right)}.
 \end{aligned}
\end{equation}
 
The coordinates of the antenna element positions are defined
\begin{equation}
\label{eq:2.12}
\begin{aligned}
x_{i} = r \sin(\theta),\\*
y_{i} = 0,\\*
z_{i} = r \cos(\theta).
\end{aligned}
\end{equation}

For the parameters $\theta$ and $\phi$, the following conditions are specified

\begin{equation}
\label{eq:2.13}
\begin{aligned}
\theta\in \left[ 0,\pi \right],\\*
\phi\in \left[ 0,2 \pi \right].
\end{aligned}
\end{equation}

In this case, the antenna axis and radiation direction are determined by the angle $\theta$. The angle $\phi$ defines all azimuthal directions of radiation.

\subsection{Numerical methods and software for modelling the brightness temperature measurement process of a microwave radiometer}\label{Software}

Numerical methods implemented in custom software written in C++ were used to model the process of measuring the brightness temperature of biological tissues with a microwave radiometer. To achieve high performance and reduce computation time, parallel computing technologies using NVIDIA CUDA were employed.

The heat conduction equation, describing the temperature distribution in biological tissues, was solved using the finite element method. The modelling domain was discretized into a tetrahedral mesh to account for the multilayered geometry of biological tissues. The main steps of the implemented finite element method are as follows:

1. The multilayer structure of biological tissue was discretised with attention to the boundaries between layers to accurately reflect differences in thermal properties.

2. The heat conduction equation was reformulated into its variational form. Local stiffness matrices and load vectors were calculated for each element and assembled into a global system of equations.

3. Convective heat exchange conditions were applied at the boundary between the biological tissue and the surrounding environment to account for thermal interaction with the environment.

4. The resulting system of equations was solved numerically using iterative methods optimised for sparse matrices, such as the conjugate gradient method. The Eigen library was used to handle sparse linear systems. Calculations were accelerated with CUDA, enabling parallel computation on an RTX 4000 GPU.

The spatial distribution of the electromagnetic field, described by the Helmholtz equation, was modelled using the finite difference method. This approach involves discretisation the domain into a uniform grid and approximating derivatives with finite differences. The Helmholtz equation was expressed in its discrete form. The Laplacian and terms containing dielectric permittivity were approximated with finite-difference expressions. The resulting discretised equation was formulated in matrix form, with each row corresponding to the equation for a single grid point. Boundary conditions specifying the absence of an electromagnetic field outside the modelling domain were applied to the tissue surface. The system was solved using the spectral decomposition method to enhance stability against high condition numbers. CUDA-based parallel computations were also applied to speed up the solution process.

The integral expression for brightness temperature, represented as a double integral over the frequency range and the volume of biological tissue, was evaluated using Simpson’s method. The frequency range and tissue volume were divided into uniform segments. For each segment, the values of the integrand were computed, and weighted sums corresponding to Simpson’s rule were calculated. The integrals for each tissue layer were computed independently, allowing efficient use of GPU resources.

Numerical results were validated using test cases with known analytical solutions.

\section{Results} \label{Results}

\subsection{Assessment of the influence of the antenna mismatch parameter on brightness temperature}

\begin{figure}[h!]
    \begin{center}
    \includegraphics[scale=1.7]{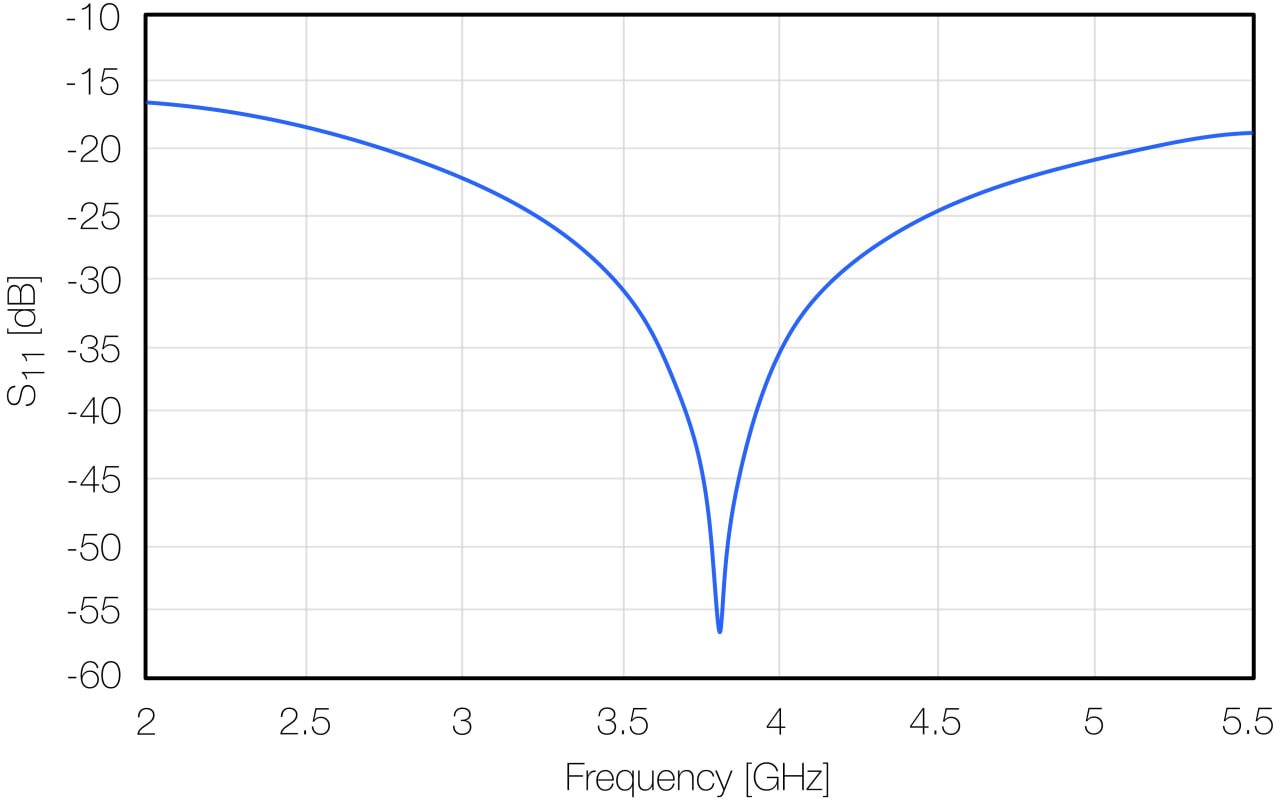}\\[2mm]
    \caption{Distribution of the $S_{11}$ parameter in the frequency range 2 GHz $\leq f \leq$ 5.5 GHz} \label{fig4}
    \end{center}
\end{figure}

To assess the impact of the antenna mismatch parameter $S_{11}$ on the brightness temperature, 5 models with different physical parameters of the internal medium were used. Figure \ref{fig4} shows the dependence of the reflection coefficient $S_{11}$ on frequency. The graph demonstrates that as the frequency increases from 2 to 5.5 GHz, the $S_{11}$ coefficient initially decreases, reaching a minimum value of about -55 dB at a frequency of around 3.8 GHz, and then increases again.
\newpage

Taking the $S_{11}$ parameter into account when calculating the brightness temperature makes a significant contribution to the final result. The experiments showed that for the 5 models presented, taking into account the parameter $S_{11}$, the brightness temperature decreased by 0.7 $^\circ$C (Table \ref{table_result}). Since the base line error of the RTM-01-RES microwave radiometer is $\pm$ 0.2 $^\circ$C, taking into account the parameter $S_{11}$ when modelling the operation of the microwave radiometer is of great importance

\begin{table}[h!] 
\caption{Brightness temperature $T_B$ without considering the parameter $S_{11}$ and brightness temperature $T_{B}^*$ considering the parameter $S_{11}$ for 5 different models} \label{table_result} \vspace{2mm}
\begin{tabular}{|c|c|c|c|c|c|}
\hline
 & Model 1 & Model 2 & Model 3 & Model 4 & Model 5 \\ \hline
$T_B$                               & 33.5     & 34.2     & 34.8     & 35.6     & 36.4     \\ \hline
$T_{B}^*$                               & 32.8     & 33.5     & 34.1     & 34.9     & 35.7     \\ \hline
\end{tabular}
\end{table}

Impedance mismatch of the antenna leads to the reflection of a portion of the electromagnetic energy, which alters the intensity of the object's radiation and, consequently, its brightness temperature. If the antenna does not match the operational conditions (such as impedance), it may cause part of the microwave radiation to reflect back to the object, which will change its brightness temperature. This can affect the accuracy of brightness temperature measurements and requires correction during data processing.

\subsection{The influence of antenna diameter on brightness temperature}

\begin{figure}[h]
    \begin{center}
    \includegraphics[scale=0.8]{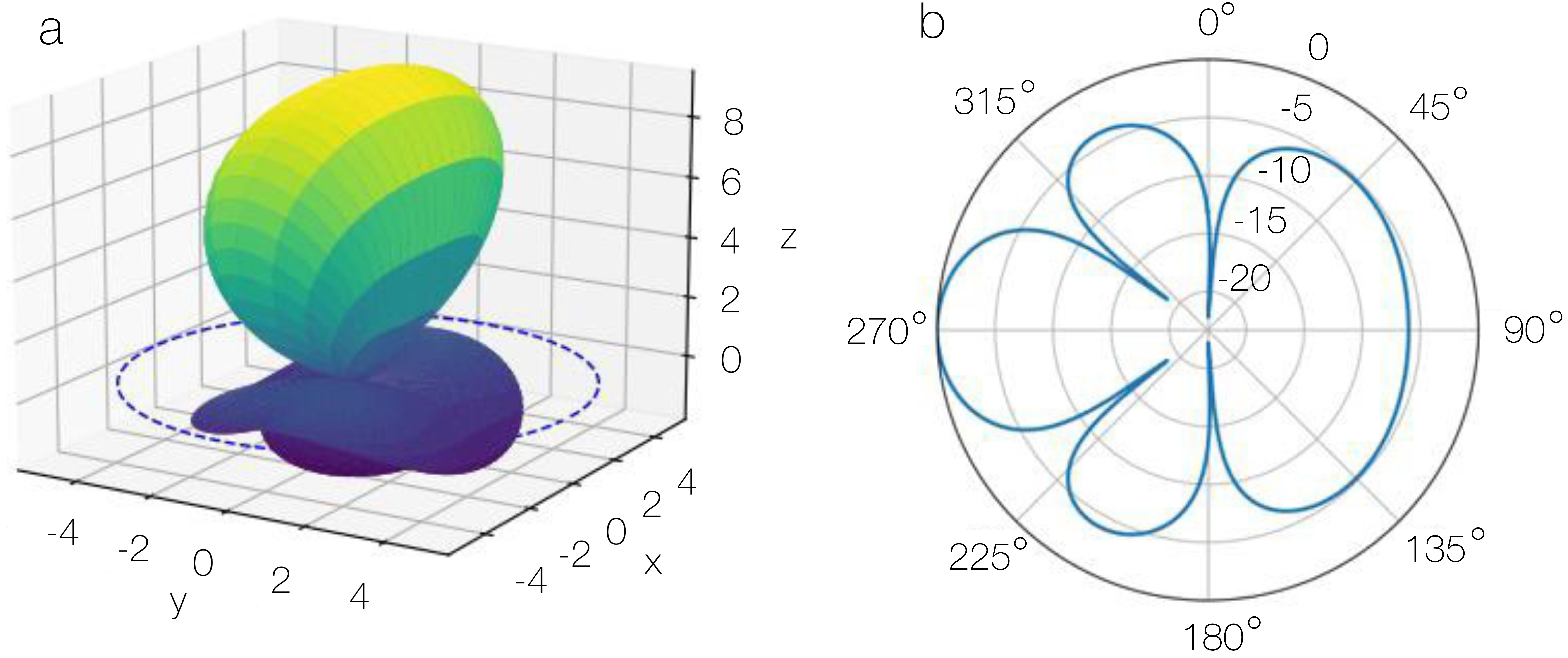}\\[2mm]
    \caption{Radiation pattern of the microwave radiometer antenna: spatial distribution of the amplitude of the radiation pattern (a), radiation pattern in polar coordinates (b)} \label{fig5}
    \end{center}
\end{figure}

Figure \ref{fig5} shows the radiation patterns of the microwave radiometer. The three-dimensional plot illustrates the spatial distribution of the amplitude of the radiation pattern. The plot demonstrates a symmetrical energy distribution, with the maximum amplitude directed along the main axis. The polar plot represents a two-dimensional section of the radiation pattern in polar coordinates. It is evident that the main lobed structure of the pattern includes several directed maxima (main and side lobes). The maximum value of the generated signal is observed along the specified direction.
\newpage

\begin{figure}[h!]
    \begin{center}
    \includegraphics[scale=0.5]{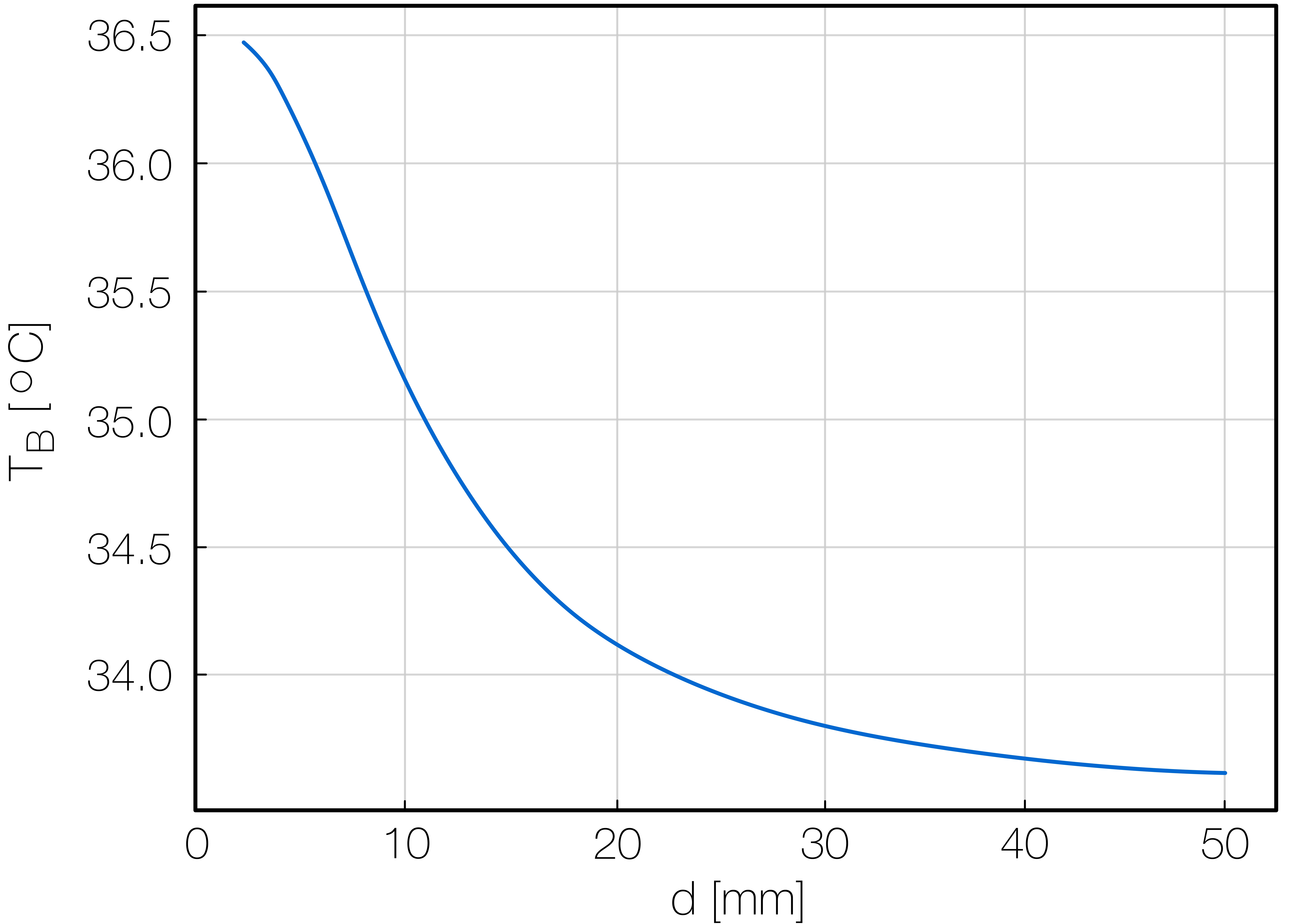}\\[2mm]
    \caption{Dependence of the brightness temperature $T_{B}$ on the diameter $d$ of the working surface of the microwave radiometer antenna} \label{fig6}
    \end{center}
\end{figure}

Figure \ref{fig6} shows the dependence of brightness temperature on antenna diameter for a microwave radiometer. The graph demonstrates a decreasing non-linear relationship. As the antenna diameter increases, the brightness temperature decreases. In the initial section (d$<$10 mm), the brightness temperature drops most rapidly, which is related to the increased sensitivity of the radiometer to the antenna size. For d$>$30 mm, the dependence stabilizes, and changes in brightness temperature become minimal, approaching an asymptotic value.

Decreasing the diameter increases sensitivity to radiation from deeper layers of the object, which leads to an increase in the brightness temperature. On the other hand, increasing the antenna diameter makes the device less sensitive to radiation from deeper layers, resulting in a lower brightness temperature.

\subsection{Influence of antenna frequency on brightness temperature}

The frequency of the radiometer antenna has a significant impact on the measured brightness temperature, as it determines the depth of penetration of electromagnetic radiation into biological tissues and the sensitivity of the method to different tissue layers.

\begin{figure}[h!]
    \begin{center}
    \includegraphics[scale=0.5]{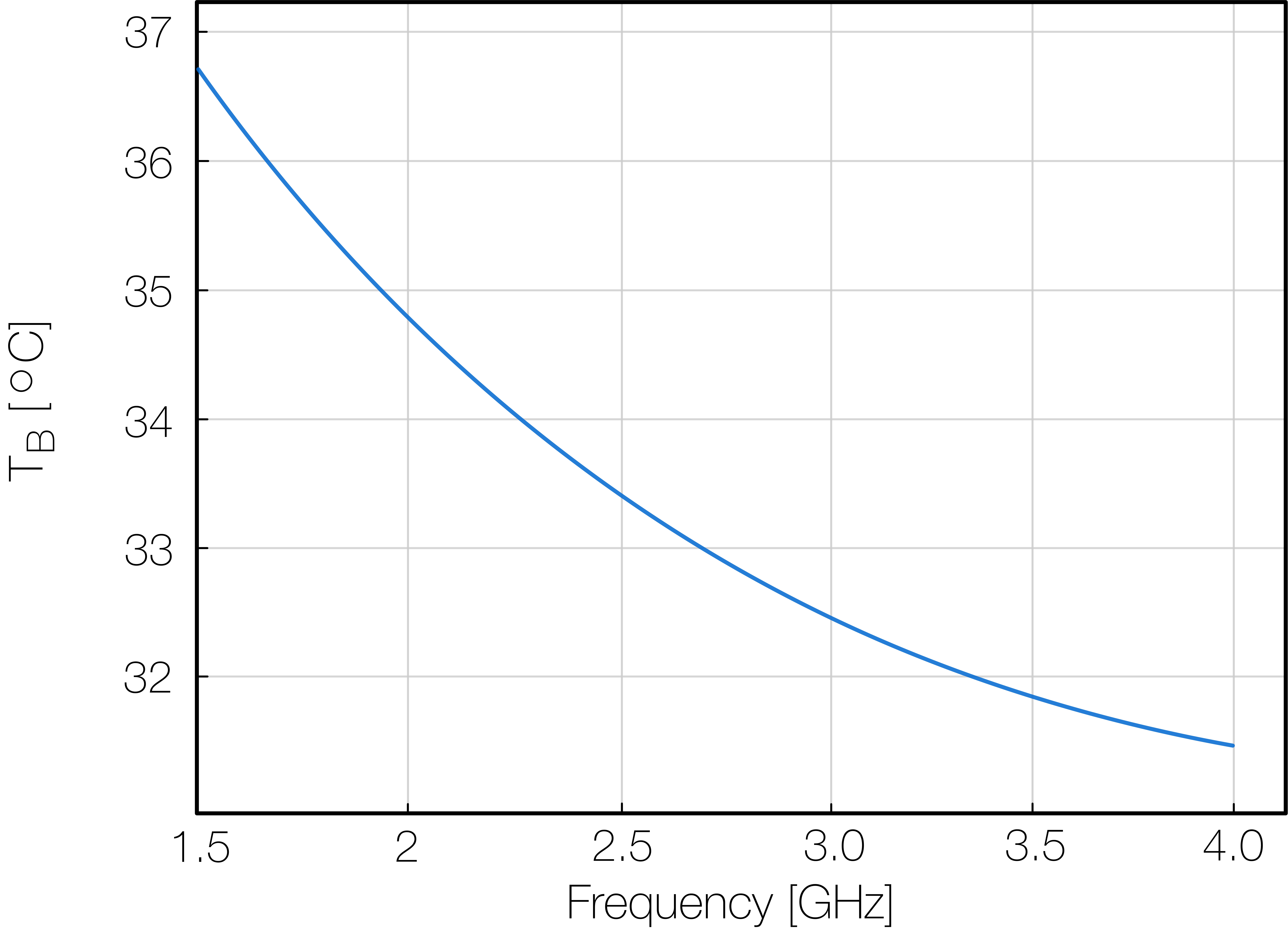}\\[2mm]
    \caption{Dependency of brightness temperature $T_{B}$ on the antenna frequency $f$ of the microwave radiometer} \label{fig7}
    \end{center}
\end{figure}




The lower the frequency of electromagnetic radiation, the longer the wavelength, and the deeper the radiation penetrates the tissues. In this case, the brightness temperature measured by the radiometer will be predominantly formed by signals coming from deeper layers. At higher frequencies, the penetration depth decreases, and the contribution to the formation of the brightness temperature comes mainly from radiation from the surface tissues (Figure \ref{fig7}).

Biological tissues have frequency-dependent dielectric properties, such as conductivity and dielectric permittivity \cite{Sharma2023}. At higher frequencies, absorption and scattering of electromagnetic radiation are enhanced, which reduces the contribution of deeper structures to the measured brightness temperature. This can reduce the method's sensitivity to changes in deeper tissues.

The brightness temperature recorded by the radiometer is an integral parameter that depends on the temperature distribution with depth and the antenna's frequency spectrum. At low frequencies, the signal from deeper, relatively warmer tissues predominates, while at higher frequencies, the brightness temperature is more influenced by the surface, often colder, layers.

\begin{figure}[h!]
    \begin{center}
    \includegraphics[scale=0.5]{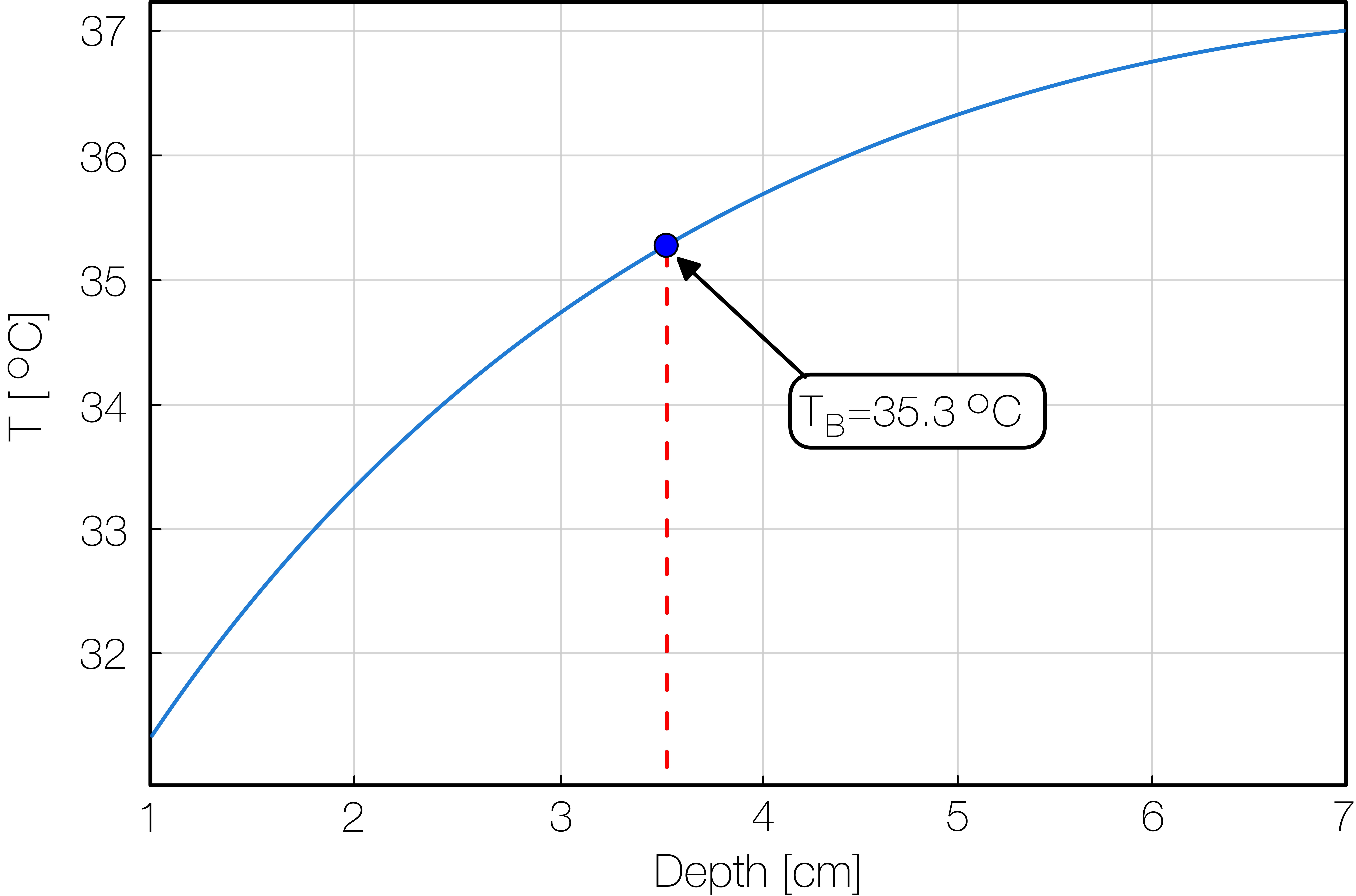}\\[2mm]
     \caption{Dependence of thermodynamic temperature on depth. The point shows the correspondence between brightness temperature and thermodynamic temperature for this model.} \label{fig8}
    \end{center}
\end{figure}

Figure \ref{fig8} shows that the brightness temperature for a multilayer model of biological tissue is formed at a depth of approximately 3.5 cm.

Unlike thermodynamic temperature, which is determined locally within each layer, brightness temperature is an integral parameter that depends on the weighted contribution of all tissue layers. The weights are determined by the radiation attenuation factor at each level. At a depth of 3.5 cm, the maximum contribution to the formation of brightness temperature is observed. This is because, at this depth, the absorption and scattering of radiation by the upper layers are balanced with the energy contribution emitted by deeper layers.

\section{Discussion and Conclusion} \label{concl}




The conducted study established that the parameters of the antenna in a medical microwave radiometer significantly influence the accuracy and reliability of tissue brightness temperature measurements.

The operating frequency of the antenna was found to be critically important for ensuring measurement precision. Experiments showed that antennas operating in the frequency range from 1 to 3 GHz delivered better results compared to higher-frequency antennas. This is because lower frequencies penetrate tissues more effectively and minimize scattering and reflection effects, reducing the likelihood of measurement distortions.

When selecting the frequency, it is also essential to consider background interference, as higher frequencies may be subject to greater noise levels, affecting measurement reliability.

This study comprehensively investigated the effect of antenna parameters of a medical microwave radiometer on the brightness temperature of biological tissues. The main conclusions are as follows:

1. A numerical model was developed to calculate the brightness temperature of biological tissues, considering key characteristics of the microwave radiometer antenna, such as frequency, matching parameters, and dimensions. The model takes into account the influence of these parameters on the depth of penetration of electromagnetic radiation into the tissue and the distribution of radiation intensity, which allows a more accurate calculation of the brightness temperature formed by radiation from different tissue layers.

2. Including the $S_{11}$ parameter in the modelling of microwave radiometers is critically important for improving measurement accuracy, as antenna mismatch causes changes in brightness temperature that exceed the device's baseline error.

3. The dependence of brightness temperature on the diameter of the microwave radiometer antenna shows that reducing the diameter increases the instrument's sensitivity to radiation from deeper tissue layers, whereas increasing the diameter decreases this sensitivity.

4. The frequency of the microwave radiometer antenna significantly affects the measured brightness temperature, determining the penetration depth of radiation and the sensitivity of the method. Lower frequencies provide a greater contribution from deep tissues, while higher frequencies are more sensitive to surface layers. This should be considered when selecting the frequency range for specific measurement tasks.

5. For multilayer biological tissues, the brightness temperature is formed at a depth of about 3.5 cm, reflecting a balance between the contributions of emissions from surface and deep tissue layers.

Studies \cite{Yazdandoost2021,Villa2023} noted that antenna frequency and size significantly influence radiation penetration depth and measurement accuracy. Our work refines this conclusion numerically, showing that increasing the antenna diameter decreases sensitivity to deep layers, while decreasing the diameter increases the contribution from deep tissue radiation. These findings complement the results of \cite{Sedankin2023}.

In \cite{Tchafa2018}, the authors emphasised that antenna mismatch significantly impacts measurement errors. However, they lacked numerical models to quantify this effect. Our work numerically demonstrates for the first time that the impact of the $S_{11}$ parameter can lead to significant deviations in brightness temperature, exceeding errors caused by receiver noise.

According to \cite{Kwon2018}, brightness temperature is formed at a depth of several centimetres, with the contribution of different layers depending on the frequency. Our work confirms and extends this finding, showing that for multilayer tissue, temperature formation occurs at a depth of about 3.5 cm, aligning with predicted values and expanding them for a specific frequency range.

In \cite{Figueiredo2018}, simplified homogeneous models of biological tissues were used to analyse infrared temperature. In contrast, our work considers multilayer structure and heterogeneity of properties, resulting in more accurate findings for complex biological objects.

Overall, this study highlights the importance of thorough analysis and optimisation of radiometer antenna parameters to ensure high accuracy and reliability of brightness temperature measurements in medical applications. Future research could focus on a deeper understanding of the interrelations between antenna parameters and object characteristics, as well as the development of new antenna optimisation methods to enhance their accuracy, including using artificial intelligence algorithms.

\vspace{2mm}

\noindent \textbf{Funding.} This work supported by the Russian Science Foundation (grant no. 23-71-00016, \href{https://rscf.ru/project/23-71-00016/}{https://rscf.ru/project/23-71-00016/}The research also relied on the shared research facilities of the HPC computing resources at Lomonosov Moscow State University.

\end{document}